\documentclass[prb,twocolumn,showpacs,superscriptaddress]{revtex4}
\usepackage{graphicx}
\usepackage{subfigure}
\usepackage{amsmath}
\usepackage{amssymb}

\begin{document}

\title{Low temperature hopping magnetotransport in paramagnetic single crystals of cobalt doped ZnO}

\author{N. Sharma}
  \affiliation{Thin Film Laboratory, Department of Physics, Indian Institute of Technology Delhi,
  Hauz Khas, New Delhi - 110016, India}
\author{S. Granville}
  \affiliation{Ecole Polytechnique F\'{e}d\'{e}rale de Lausanne, ICMP, Station 3, CH-1015 Lausanne-EPFL, Switzerland}
\author{S. C. Kashyap}
  \affiliation{Thin Film Laboratory, Department of Physics, Indian Institute of Technology Delhi,
  Hauz Khas, New Delhi - 110016, India}
\author{J.-Ph. Ansermet}
  \affiliation{Ecole Polytechnique F\'{e}d\'{e}rale de Lausanne, ICMP, Station 3, CH-1015 Lausanne-EPFL, Switzerland}

\begin{abstract}
Long needle-shaped single crystals of Zn$_{1-x}$Co$_{x}$O were grown at low temperatures using a molten salt solvent technique, up to $x=0.10$. The conduction process at low temperatures is determined to be by Mott variable range hopping. Both pristine and cobalt doped crystals clearly exhibit a crossover from negative to positive magnetoresistance as the temperature is decreased. The positive magnetoresistance of the Zn$_{1-x}$Co$_{x}$O single crystals increases with increased Co concentration and reaches up to 20~\% at low temperatures (2.5~K) and high fields ($>$1~T). SQUID magnetometry confirms that the Zn$_{1-x}$Co$_{x}$O crystals are predominantly paramagnetic in nature and the magnetic response is independent of Co concentration. The results indicate that cobalt doping of single crystalline ZnO introduces localized electronic states and isolated Co$^{2+}$ ions into the host matrix, but that the magnetotransport and magnetic properties are decoupled.
\end{abstract}


\date{\today}

\maketitle

\section{Introduction}

Due to a predicted high Curie temperature (above 300~K), diluted magnetic semiconductors (DMS) are regarded as important materials for spintronics.~\cite{Dietl_Ferrand} Combining ferromagnetism with semiconductivity in oxides gives them an additional degree of freedom and functionality for fabricating unique devices with applications ranging from nonvolatile memory to quantum computing.~\cite{Ohno,Wolf} Some theoretical~\cite{Katayama-Yoshida} and experimental~\cite{Ueda_Kawai,Prellier_Raveau,Kim_Park} reports have shown room temperature ferromagnetism (RTFM) in highly doped ZnO films. The magnetic impurities are either implanted into the host matrix~\cite{Norton_Wilson} or introduced during the growth.~\cite{Schwartz_Gamelin} The experimental results regarding RTFM and its origin in Zn$_{1-x}$Co$_{x}$O samples in different forms including thin films, bulk, polycrystalline and nanocrystalline samples differ widely, and there is no consensus on either the nature of the magnetism (whether the samples are para-, dia- or ferro-magnetic) or on the origin of the ferromagnetism (intrinsic, extrinsic, defect induced etc.).

In the case of cobalt doped ZnO films, the observed RTFM has been attributed either to Co clustering,~\cite{Kim_Choo} some external pollution~\cite{Colis_Dinia} or is thought to be intrinsic in nature.~\cite{Sati_Golacki} Cobalt and lithium co-doped ZnO films have revealed only Co paramagnetism, and oxygen vacancies were considered as the origin for the RTFM.~\cite{Tietze_Goering} In bulk samples the FM is attributed to Co clustering,~\cite{Deka_Joy1} and in hydrogenated samples it is attributed either to the appearance of Co clusters~\cite{Deka_Joy2} or oxygen vacancies.~\cite{Manivannan_Seehra} In the case of polycrystalline and nanocrystalline samples, some groups have reported paramagnetism.~\cite{Kolesnik_Mais,Yin_Jiang,Lawes_Seshadri} Recently, nanocrystals of Zn$_{1-x}$Co$_{x}$O have been reported to be ferromagnetic~\cite{Maensiri_Phokha} or superparamagnetic,~\cite{Martinez_Monty} and nanowires showed a paramagnetic behavior with Co concentrations up to $x=0.05$.~\cite{Enculescu_Ansermet} Very recently, thin films and nanoparticles of pure ZnO have been reported to be ferromagnetic, and their FM is attributed to defects on Zn sites and surface point defects.~\cite{Hong_Brize,Sundaresan_Rao}

From a practical point of view, the ultimate goal of investigating materials such as cobalt doped ZnO is to fabricate semiconductor devices that have a spin-polarized nature to their electronic transport properties. To this end, explorations of the magnetotransport properties are an essential part of studies into cobalt doped ZnO.  The low temperature magnetoresistance (MR) of cobalt doped ZnO is typically tens of percent in magnetic fields of up to several Tesla,~\cite{Tian_Yan} and may show a crossover between positive and negative MR~\cite{Gacic_Goering,Dietl_Wu,Jin_Kawasaki} at different temperature and magnetic field values. ZnO films containing magnetic nanoclusters within a non-magnetic insulating matrix have positive MR values as large as 811\% at 5~K in fields of a few Tesla,~\cite{Tian2} revealing the rich potential for magnetic-field influenced charge transport in DMS.

The exploration of single crystals of DMS is of great significance as this form is ideally suited for understanding the intrinsic properties of any material. In the present paper we present the structural, magnetotransport and magnetic properties of both pristine ZnO and cobalt doped ZnO single crystals, which were grown by the molten salt solvent technique~\cite{Kashyap}.

\section{Experimental details}

In the present case, the molten salt solvent technique has been employed to grow single crystals of both pristine ZnO and cobalt doped ZnO. Analytical Reagent (AR) grade ZnO, KOH$\cdot$6H$_{2}$O and CoCl$_{2}\cdot$2H$_{2}$O were used as precursors. Calculated amounts~\cite{Kashyap} of ZnO and CoCl$_{2}\cdot$2H$_{2}$O (an appropriate amount was chosen corresponding to the desired concentration of Co) were slowly added to the molten KOH. The crucible was then covered with a specially designed silver lid which was fitted with a silver tube at its centre. The other end of the silver tube was tapered into a cone. The crucible with the charge (solvent + solute) was then placed in the furnace such that the charge remained in the uniform temperature region of the furnace at a temperature of 480$^{\circ}$C for $\sim$50~hours. When immersed in the molten charge, the conical end of the tube provides suitable sites for the growth of the crystals, allows the heat of solidification to escape and also serves as a sheath for the thermocouple (one end of which is placed in the tube). These conditions result in the nucleation and isothermal growth of transparent, green-colored, needle-shaped cobalt doped single crystals of ZnO due to dissociation of zincate species unstable at 480$^{\circ}$C, at the cylindrical walls of the silver crucible and the tapered end of the silver tube.~\cite{Kashyap} The crystals were extracted by dissolving the solvent in methyl alcohol. Dimensions of the needle-like crystals are typically 2.5 to 5~mm in length, 0.5 - 1.0~mm in width and 0.1 to 0.2~mm in thickness.

The single crystals of pristine and cobalt doped ZnO were powdered and their structural properties were studied using a Philips (X'Pert PRO, Model PW 3040) x-ray diffraction (XRD) system with Cu K$_{\alpha}$ radiation at wavelength 1.54060~{\AA}. The SEMCF used for SEM micrographs has RONTECs EDAX (Energy Dispersive Analysis of X-rays) system Model QuanTax 200, which is based on Silicon Drift Detector (SDD) technology and provides an energy resolution of 127~eV at Mn K$_{\alpha}$. The magnetic moment of a small amount of each kind of crystal was measured as a function of both applied magnetic field and temperature, using a Quantum Design MPMS-7 SQUID magnetometer. The temperature and magnetic field dependent electrical resistance of individual crystals was also measured. The maximum applied magnetic field for these measurements was 4.5 T, and the range of temperatures employed was 2.5 - 350~K.

\section{Results and Discussion}

\subsection{Structure and composition}
\begin{figure}\vspace{-0.4 cm}
\resizebox{\columnwidth}{!} {\includegraphics{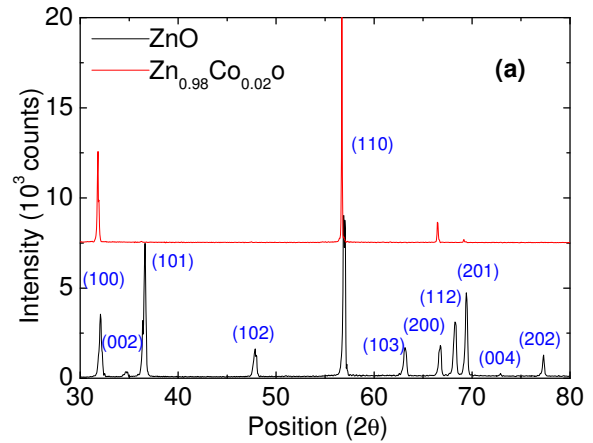}}
\resizebox{\columnwidth}{!} {\includegraphics{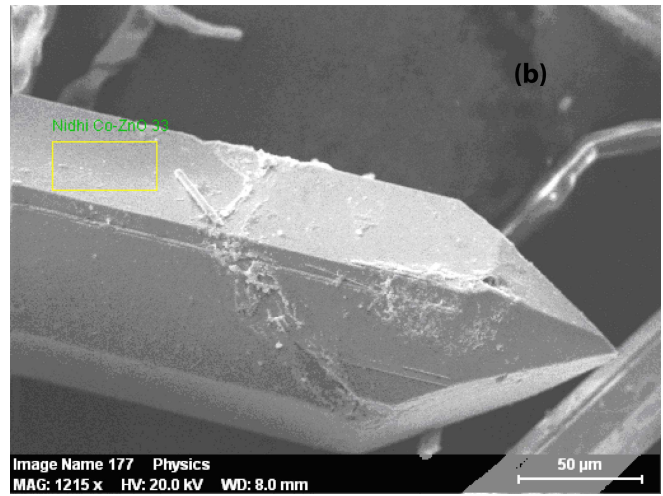}}
\caption{(Color online) (a) Powder x-ray diffractograms of crystals of pristine ZnO and Zn$_{0.98}$Co$_{0.02}$O. The peaks are labeled with the reflections corresponding to the ZnO wurtzite structure. (b) Scanning electron micrograph of Zn$_{0.98}$Co$_{0.02}$O single crystals.
}
\label{fig1}
\end{figure}

The x-ray diffractograms of both the pristine ZnO and Zn$_{1-x}$Co$_{x}$O crystals are shown in Fig.~\ref{fig1}(a). All the diffraction peaks correspond to ZnO in the wurtzite structure. No secondary phases are detected in the XRD spectrum of Zn$_{1-x}$Co$_{x}$O crystals, which implies that the Co$^{2+}$ ions are incorporated into the lattice substitutionally replacing Zn$^{2+}$. It can also be seen that the XRD peaks of the Zn$_{1-x}$Co$_{x}$O are slightly shifted to lower angles as compared to the pure ZnO peaks, which implies that the lattice is slightly expanded relative to pure ZnO.
                                                                                                                       					
Fig.~\ref{fig1}(b) shows the morphology of a Zn$_{1-x}$Co$_{x}$O single crystal as studied using scanning electron microscopy. This depicts the characteristic hexagonal shape of a ZnO single crystal. The exposed faces are hexagonal \textbf{m} {10$\overline{1}$0} and hexagonal \textbf{p} {10$\overline{1}$1} (cone) faces. The precise elemental composition was calculated from the EDAX spectrum; the concentration of Co in the two samples has been estimated to be $x=0.02$ and $x=0.10$.

\begin{figure} \vspace{-0.4 cm}
\resizebox{\columnwidth}{!} {\includegraphics{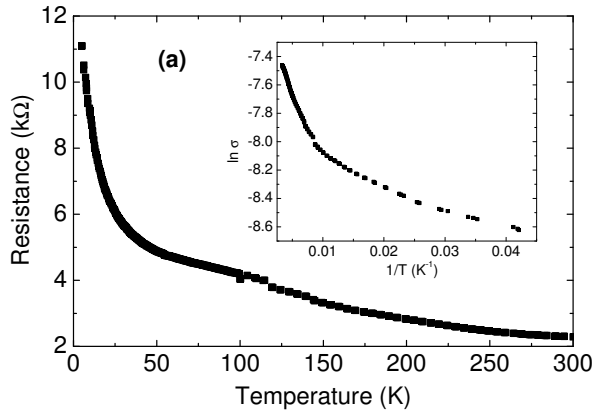}} \vspace{-0.4 cm}
\resizebox{\columnwidth}{!} {\includegraphics{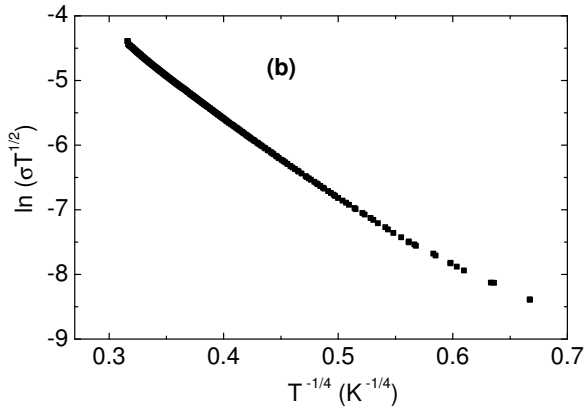}}
\caption{(a) Resistance vs temperature curve for a Zn$_{0.90}$Co$_{0.10}$O single crystal. Inset: Plot of $\ln \sigma$ vs 1/T, indicating that the conduction mechanisms are different at high and low temperatures. (b) Plot of $\ln \left(\sigma T^{1/2}\right)$ vs $T^{-1/4}$ for T$<$100~K indicating that the variable range hopping conduction mechanism is active in this temperature range (data shown for $x=0.10$).}
\label{fig2}
\end{figure}

\subsection{Electronic properties - temperature and field dependent resistance}

\begin{table*}\hspace{-6.0 cm}
\caption{\label{table1}Parameters derived from fitting the variable range hopping model of Eqs.~\ref{MottVRH} and~\ref{MottVRHcoeff1} to the temperature dependent resistivity data of the ZnO and Zn$_{1-x}$Co$_x$O crystals. $T$ is the temperature used in calculating parameters $R$ and $W$ as in Eq.~\ref{MottVRHparams}. All other parameters are as detailed in the text.}
\begin{ruledtabular}
\begin{tabular}{lccccccccccc}
Sample & Intercept & Gradient & $\sigma_{h0}T^{\frac{1}{2}}$ & $T_{0}$ & $\alpha$ & $N(E_F)$ & $R$ & $T$ & $W$ & $\alpha$$R$ \\
 & & & (Scm$^{-1}$K$^{\frac{1}{2}}$) & (K) & (cm$^{-1}$) & (eV$^{-1}$cm$^{-3}$) & (cm) & (K) & (meV) & & \\
\hline
$x=0.02$ & 6.71 & -13.59 & 820.57 & 34109.7 & 3.41$\times10^6$ &$2.161\times10^{20}$ &$6.89\times10^{-7}$ & 25 & 3.4 & 2.35 \\
$x=0.10$ & 7.53 & -9.35 & 1863.1 & 7642.7 & $3.67\times10^6$ & $1.20\times10^{21}$ & $4.41\times10^{-7}$ & 25 & 2.3 & 1.62 \\
ZnO 25-100~K & 7.24 & -5.88 & 1394.09 & 1195.4 & $1.09\times10^6$ & $1.98\times10^{20}$ & $9.37\times10^{-7}$ & 25 & 1.5 & 1.02 \\
ZnO 10-60~K & 6.65 & 4.41 & 772.78 & 378.2 & $3.38\times10^5$ & $1.90\times10^{19}$ & $2.83\times10^{-6}$ & 10 & 0.56 & 0.96 \\
\end{tabular}
\end{ruledtabular}
\end{table*}

To study the electronic transport properties of the crystals, the temperature dependent resistance behavior was observed for Zn$_{1-x}$Co$_{x}$O single crystals. Electrical contacts were made to individual crystals using silver paste.  The 300~K resistivity of the pristine ZnO crystals is 3.6~$\Omega$cm, and the Zn$_{1-x}$Co$_x$O crystals 5.4~$\Omega$cm and 6.7~$\Omega$cm with $x$=0.10 and $x$=0.02 respectively. The exponential decrease of resistance with increase in temperature as shown in Fig.~\ref{fig2}(a) is a clear signature that the cobalt doped ZnO maintains the semiconducting behavior of undoped ZnO. The two different slopes of the ln~$\sigma$ vs 1/T (inset of Fig.~\ref{fig2}(a)) curve in the low (2.5-100~K) and high temperature (100-300~K) regions are signatures of two different conduction mechanisms being operative in these two temperature regions. At high temperatures, the conduction mechanism is thermally activated band conduction from donor levels near the conduction band, with an activation energy of 10~meV.~\cite{Kumar_Khare} At low temperatures there is insufficient thermal energy to promote conduction through delocalized carriers. Instead, below we show that the variable range hopping (VRH) conduction mechanism is active at low temperatures, where the electrons hop between levels that are close to the Fermi level irrespective of their spatial separation. The relation between electrical conductivity and temperature for VRH is given by Mott's equation~\cite{Mott}

\begin{equation}\label{MottVRH}
\sigma=\sigma_{h0}\exp^{-(T_{0}/T)^p}.
\end{equation}

\noindent The parameters $\sigma_{h0}$ and $T_{0}$ are given by the following expressions~\cite{Paasch_Scheinert,Brodsky}

\begin{equation}\label{MottVRHcoeff1}
\sigma_{h0}=3e^2\nu_{ph}\left(\frac{N(E_F)}{ 8\pi\alpha k_BT}\right)^{1/2};  T_0=\frac{16\alpha^3}{k_BN(E_F)},
\end{equation}

\noindent where $\nu_{ph}$ is the phonon frequency at the Debye temperature, $k_{B}$ is Boltzmann's constant, $N(E_{F})$ is the density of localized electronic states at the Fermi level and $\alpha$ is the inverse localization length. The exponent $p$ in Eq.~\ref{MottVRH} varies from $p\sim\frac{1}{4}$ when the density of states across the Fermi level is constant to $p\sim\frac{1}{2}$ in the case of significant electron-electron interactions that lead to the formation of a Coulomb gap. From Eqs.~\ref{MottVRH} and~\ref{MottVRHcoeff1} we expect~$\ln(\sigma$$T^{1/2}) \propto T^{-1/p}$. The nearly linear plot of~$\ln(\sigma$$T^{1/2})$ vs~$T^{-1/4}$ in Fig.~\ref{fig2}(b) indicates that VRH is the dominant mechanism of conduction in Zn$_{0.90}$Co$_{0.10}$O below 100~K.  We find similar behavior for both undoped ZnO and Zn$_{0.98}$Co$_{0.02}$O crystals. The results of fitting the data for all three samples with the VRH model (Eqs.~\ref{MottVRH} and~\ref{MottVRHcoeff1}) are reported in Table~\ref{table1}. The Mott parameters, i.e. average hopping distance ($R$) and average hopping energy ($W$), were calculated from

\begin{equation}\label{MottVRHparams}
R=\frac{9}{8\pi k_BTN(E_F)};  W=\frac{4}{3\pi R^3N(E_F)}.
\end{equation}

\noindent Other conditions for VRH conduction are that the value of $\alpha$$R$ must be greater than~1, and that $W>k_BT$.  Note that both of these conditions are satisfied in the Zn$_{1-x}$Co$_x$O crystals. For the ZnO crystals the conditions are only satisifed above approximately 25~K, and at lower temperatures other conduction mechanisms may compete with VRH. The values of $T_0$, $N(E_F)$ and $R$ determined by fitting to the VRH model are reasonable compared with results which exist for both doped and undoped ZnO.~\cite{Bandyo_Paul,Singh_Rao,Kumar_Khare,Huang_Chiu} Thus, the analysis of the temperature dependent electronic transport data for both the ZnO and Zn$_{1-x}$Co$_x$O crystals provides convincing evidence for the VRH conduction process up to approximately 100~K, above which a thermally activated conduction mechanism becomes more appropriate.

\begin{figure} \vspace{-0.6 cm}
\resizebox{\columnwidth}{!} {\includegraphics{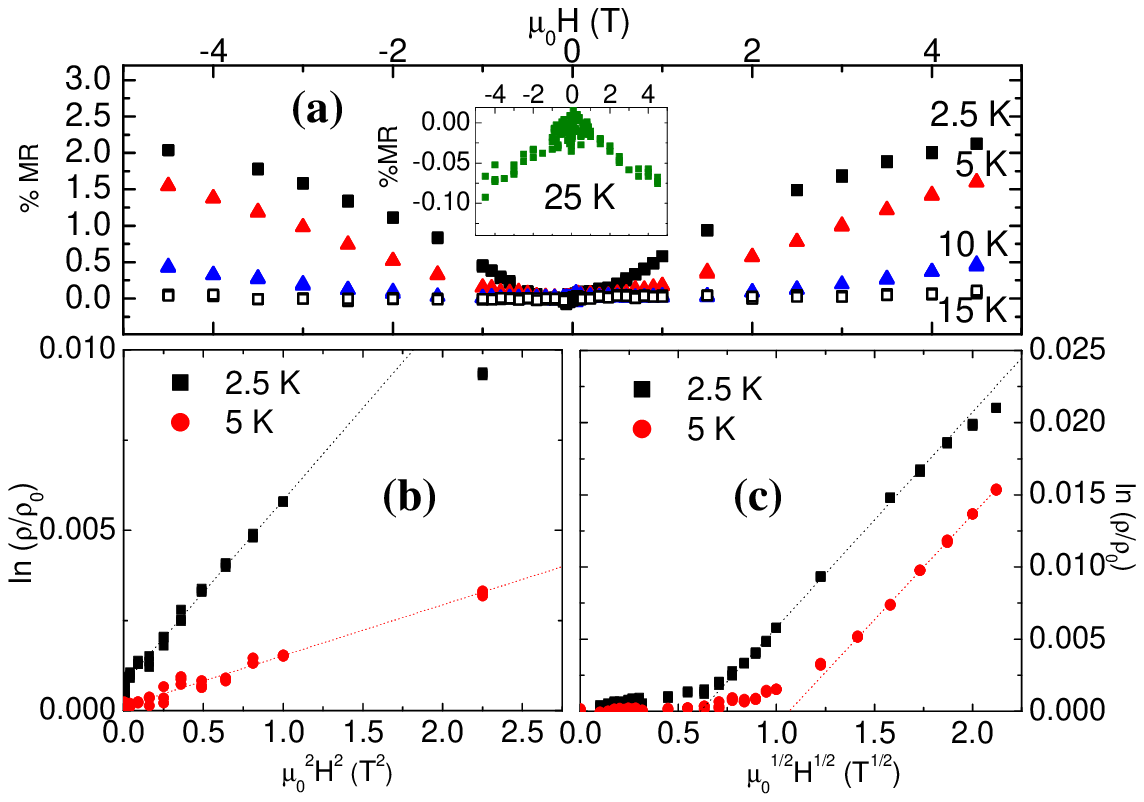}}\vspace{-0.4 cm}
\resizebox{\columnwidth}{!} {\includegraphics{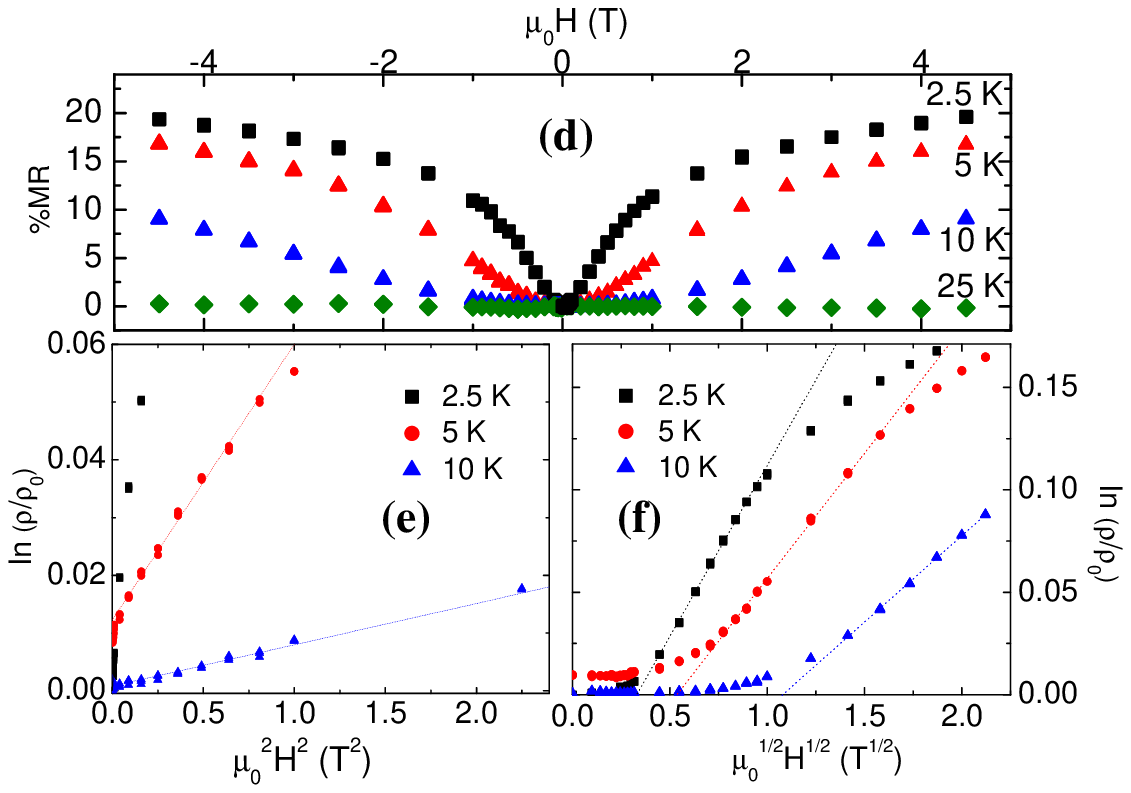}}\vspace{-0.4 cm}
\caption{(Color online) Magnetoresistance of crystals of (a) Zn$_{0.98}$Co$_{0.02}$O and (d) Zn$_{0.90}$Co$_{0.10}$O below 25~K. Inset of (a): zoomed view showing the magnetoresistance at 25~K. (b)-(c) Field dependence of ln ($\rho$/$\rho_0$) for a Zn$_{0.98}$Co$_{0.02}$O crystal, plotted versus $\mu_0^2H^2$ and $\mu_0^{1/2}H^{1/2}$. (e)-(f) Field dependence of ln ($\rho$/$\rho_0$) for a Zn$_{0.90}$Co$_{0.10}$O crystal, plotted versus $\mu_0^2H^2$ and $\mu_0^{1/2}H^{1/2}$. Dotted lines are guides to the eye.
}\label{fig3}
\end{figure}

Measurements were also carried out to study the magnetoresistive behavior of the single crystals up to a magnetic field $\mu_{0}H$ of 4.5~T and at temperatures down to 2.5~K. Figs.~\ref{fig3}(a) and~\ref{fig3}(d) show that for the Zn$_{1-x}$Co$_x$O crystals the magnetoresistance (MR) is weak and negative (see inset Fig.~\ref{fig3}(a)) down to a certain temperature. Below that temperature the MR becomes positive (i.e. changes its sign) and becomes larger in magnitude. The crossover in MR from negative to positive occurs at $\sim$20~K in the doped crystals.  It may be noted that the MR increases with the decrease in temperature in all cases, and also with increase in the concentration of cobalt. The largest value attained, near saturation of the positive MR, is 20~\% (Fig.~\ref{fig3}(d)) at 2.5~K for Zn$_{0.90}$Co$_{0.10}$O single crystals, which is at the upper end of MR values reported by other groups for ZnO-based DMS~\cite{Jin_Kawasaki,Yang_Birge,Gacic_Goering,Dietl_Wu,Venkatesan_Coey}. A positive MR also arises in pristine ZnO crystals below 5~K (not shown), but it is weaker than in the cobalt doped crystals, no more than 1\% at 2.5~K, which is likely due to residual impurities.
	
There are several mechanisms which are typically cited to account for the MR in DMS at low temperatures:
\begin{itemize}
\item a negative MR at low fields owing to the magnetic field breaking the time-reversal symmetry of scattering paths involved in weak localization;
\item a negative MR at large fields from a reduced scattering of the carriers as the magnetic field aligns the moments of the magnetic dopants;
\item a positive MR at intermediate fields from a field-induced spin-splitting of the conduction band due to the s-d exchange interaction.

\end{itemize}
\noindent The absence of evidence for the first type, i.e. a sharp negative MR at $<$0.1~T, as seen in a number of works~\cite{Jin_Kawasaki,Dietl_Wu}, is consistent with the temperature dependent resistance measurements of Fig.~\ref{fig2}, which indicate a hopping type conduction at low temperatures rather than weak localization of the carriers. The second type of MR is seen in thin films of ZnO-based DMS that exhibit ferromagnetism in magnetic measurements, and the absence of this component at the magnetic fields applied is consistent with the paramagnetic character of the single crystals as measured by SQUID magnetometry (discussed below). A phenomenological model for the third mechanism of MR has been developed by Khosla and Fischer~\cite{Khosla_Fischer}, where the positive MR is proportional to $H^2$.  This model has been used elsewhere to explain the MR behavior of doped ZnO thin films~\cite{Gacic_Goering,Liu_Jiang}. The predicted $H^2$ dependence appears to account for the data of Figs.~\ref{fig3}(b) and~\ref{fig3}(e). However, the Khosla-Fischer model considers the MR to arise from a field-induced splitting of bands of extended states. In the Zn$_{1-x}$Co$_x$O crystals of the present report, conduction is via localized states rather than through bands of extended states.  Hence, none of the mechanisms of MR commonly applied to DMS are applicable in our case, and we must consider an alternate mechanism to explain the low temperature positive MR, based on the VRH conduction process.

\begin{figure} \vspace{-0.6 cm}
\resizebox{\columnwidth}{!} {\includegraphics{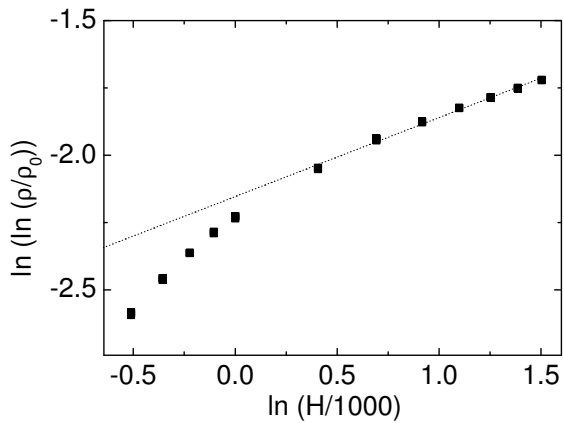}}\vspace{-0.4 cm}
\caption{Magnetoresistance of Zn$_{0.90}$Co$_{0.10}$O at 2.5~K, plotted to show that $\ln\left(\frac{\rho(H,T)}{\rho(0,T)}\right)\propto$$H^{0.3}$. The dotted line is a least-squares fit of the data for $\mu_0$$H>1.5$~T indicating a gradient of 0.3.
}\label{fig4}
\end{figure}

At low temperatures, where hopping distances are large enough, a positive MR arises due to the shrinking of the wavefunctions of the states available for an electron hop.~\cite{Shklovskii_Efros} In low magnetic fields, the positive MR can then be described with

\begin{equation}\label{MottVRHMR2}
\ln\left(\frac{\rho(H,T)}{\rho(0,T)}\right)\propto\frac{\xi^4e^2H^2}{c^2\hbar^2}\left(\frac{E_H}{k_BT}\right)^{3p} \\
\end{equation}

\noindent where $\xi$ is the carrier localization length and $p$ is the VRH exponent of Eq.~\ref{MottVRH}. This may account for the $H^2$ dependence at low fields as shown in Figs.~\ref{fig3}(b) and~\ref{fig3}(e).  At higher fields, there is a change to an approximate $H^{1/2}$ behavior, as seen before in thin films of cobalt doped ZnO in the low temperature 'hard gap' regime of electronic transport~\cite{Tian_Yan}. Note that at the lowest temperatures and highest fields, especially for the higher Co concentration crystals, the MR is more weakly field dependent than in the low field region.  The data of Fig.~\ref{fig4} show that, at 2.5~K in fields of $\mu_0H>$1.5~T, the MR of the Zn$_{0.9}$Co$_{0.1}$O crystals can be approximately described by $\ln\left(\frac{\rho(H,T)}{\rho(0,T)}\right)\propto$$H^{0.3}$, which is very close to the expected high-field MR in the VRH model of $\ln\left(\frac{\rho(H,T)}{\rho(0,T)}\right)\propto$$H^{1/3}$~\cite{Shklovskii_Efros}. We thus find that the MR properties of the Zn$_{1-x}$Co$_x$O crystals are completely explained as the effect of the magnetic field on the VRH conduction process.

\subsection{Magnetization}

\begin{figure} \vspace{-0.4 cm}
\resizebox{\columnwidth}{!} {\includegraphics{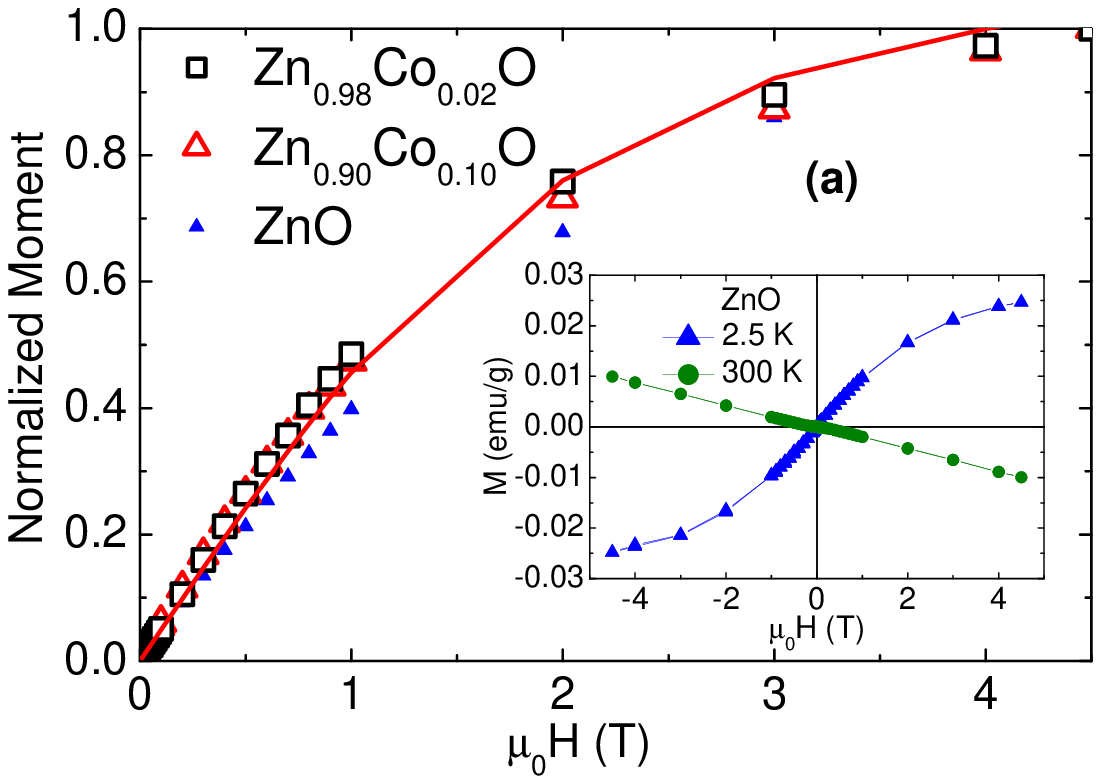}}\vspace{-0.4 cm}
\resizebox{\columnwidth}{!} {\includegraphics{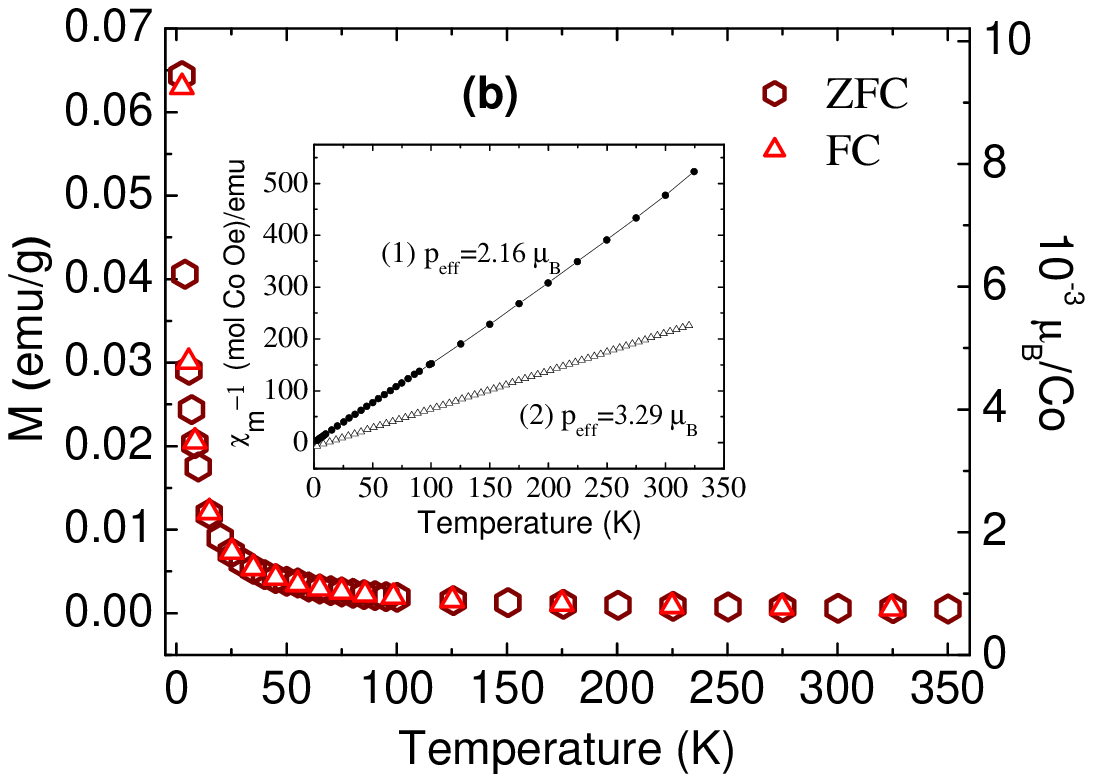}}\vspace{-0.4 cm}
\caption{(Color online) (a) Magnetization field loops for pristine and cobalt doped ZnO single crystals at 2.5~K, normalized to the value at 4.5~T (blue triangles - ZnO, black squares -  Zn$_{0.98}$Co$_{0.02}$O , red circles - Zn$_{0.90}$Co$_{0.10}$O, red line - Brillouin function for ions with S = 3/2). Inset: $M$ vs $\mu_0H$ for ZnO at 2.5~K (blue triangles) and 300~K (green circles). (b) ZFC and FC curves for Zn$_{0.90}$Co$_{0.10}$O measured in a field of 50~mT. Inset: $\chi^{-1}$ vs $T$ curve for cobalt doped ZnO crystals. Sample (1) - Zn$_{0.90}$Co$_{0.10}$O; Sample (2) - Zn$_{0.98}$Co$_{0.02}$O.
}\label{fig5}
\end{figure}

The results of magnetic measurements on each crystal type are shown in Fig.~\ref{fig5}. The magnetization vs field ($M$-$H$) measurements were carried out at room temperature as well as at 2.5~K. The $M$-$H$ curves of Fig.~\ref{fig5}(a) for cobalt doped samples at 2.5~K, normalized to their respective values at 4.5~T for each sample, exhibit neither remanence nor coercivity. The maximum values of magnetization of the Zn$_{0.98}$Co$_{0.02}$O and Zn$_{0.90}$Co$_{0.10}$O crystals in 4.5~T applied field are 2.01 and 2.34~emu/g respectively, corresponding to absolute magnetic moment values of 5$\times10^{-3}$-1$\times10^{-3}$ emu. In addition, the curves do not saturate, even at the highest applied field. A representative sample of the temperature dependent magnetization ($M$-$T$) measurements for the crystals, measured under both zero-field cooled ($M^{ZFC}(T)$) and field-cooled ($M^{FC}(T)$) conditions, is plotted in Fig.~\ref{fig5}(b). A ferromagnetic material, besides exhibiting distinct hysteresis in an $M$-$H$ loop, also demonstrates thermomagnetic irreversibility (TMI), i.e., $M^{FC}(T)\neq$$M^{ZFC}(T)$ in $M$-$T$ curves. The $M^{FC}$ and $M^{ZFC}$ curves of Fig.~\ref{fig5}(b) are identical, showing no sign of TMI, further confirming the lack of ferromagnetic ordering in the sample; instead the observed behavior is consistent with paramagnetism. It is thus evident that the Zn$_{1-x}$Co$_x$O single crystals are paramagnetic, and doping of Co does not result in any ferromagnetism.

To confirm that the magnetic behavior observed is due to the cobalt doping, the $M$-$H$ curves of the pristine ZnO sample recorded at both 2.5~K and room temperature (300~K) are drawn in the inset of Fig.~\ref{fig5}(a). It may be noted that the data recorded at room temperature exhibit a linear variation of magnetization $M$ with applied field, with a negative slope. This is a signature of a diamagnetic material. However, at 2.5~K pristine ZnO crystals reveal a feeble paramagnetism with a magnetization 100$\times$ weaker than that of the Zn$_{1-x}$Co$_x$O. This weak low-temperature paramagnetism is possibly due to the presence of residual impurities as detected in the MR measurements.

In paramagnets, a value of the effective magnetic moment per ion, $p_{eff}$, can be estimated from the slope ($=\frac{Np_{eff}^2}{3k_B}$) of a $\chi^{-1}$ vs $T$ plot (Curie law fit), where $N$ is the number of moles of Co. The inset of Fig.~\ref{fig5}(b) shows that the Curie law applies over the full range of temperatures measured. The value of $p_{eff}$ was found to be 3.29~$\mu_B$ for Zn$_{0.98}$Co$_{0.02}$O, which compares favorably to values determined for other paramagnetic ZnO:Co samples~\cite{Yin_Jiang,Zhang_Li,Lawes_Seshadri} and is close to the theoretical free Co$^{2+}$ ion spin-only value of 3.87~$\mu_B$. However, at the higher Co concentration of $x=0.10$, $p_{eff}$ has a lower value of 2.16~$\mu_B$. As the Co concentration increases, the average Co-Co distance is reduced, which strengthens the direct antiferromagnetic exchange between Co neighbors or superexchange via Co-O-Co bonds, and consequently the average moment per Co is reduced from the case of isolated non-interacting Co$^{2+}$ ions. Theoretically, the value of the total moment of paramagnetic Co$^{2+}$ ions is $J$(=$L$+$S$)=3+3/2=9/2. However, for transition metals in semiconductors such as ZnO it is often found that the orbital angular momentum of the unpaired 3$d$ electrons is quenched, i.e. $L$=0, and hence the total moment comes only from the spin momentum $S$=3/2. A Brillouin function for the field dependence of the moment of isolated, non-interacting paramagnetic Co$^{2+}$ ions with a spin-only moment of 3/2 is plotted in Fig.~\ref{fig5}(a) alongside the magnetic moment data measured at 2.5~K. The calculated field dependence for $S$=3/2 paramagnetic ions is extremely close to the experimental data from the cobalt doped samples, confirming that the field dependent magnetization is from a majority of Co ions in a divalent, paramagnetic state. The antiferromagnetically correlated fraction of Co ions in the crystals is insensitive to the magnetic field.

Finally, we summarize the state of the incorporated Co ions in the single crystals of ZnO, and their effect on the magnetic and magnetotransport properties. The resistivity and MR measurements point to the presence of localized electronic states, whereas the magnetic properties result from isolated Co$^{2+}$ ions in the crystals. At higher Co concentrations, a proportion of the Co ions experience some antiferromagnetic exchange, likely with near neighboring Co ions, which reduces the measured effective moment per incorporated ion. At low temperatures, charge carriers hop between the localized electronic states, and in the presence of a magnetic field the probability of hopping is reduced as the wavefunctions of the localized states shrink. However, at higher magnetic fields, the paramagnetic Co ions progressively align with the field, and a spin dependent VRH conduction process may be expected to lower the resistivity at sufficiently high fields~\cite{Wagner,Yan_Liu}.  No such negative MR is seen in Fig.~\ref{fig3}, even at the lowest temperature and maximum field, where the $M$-$H$ data of Fig.~\ref{fig5}(a) show that magnetic alignment of the Co ions is significant. Furthermore, the MR is evidently decoupled from the magnetization of the crystals, as demonstrated by the different shapes of the MR in the two crystals as compared to the identical shape of their $M$-$H$ curves. This independence of the magnetotransport from the magnetization in the Zn$_{1-x}$Co$_x$O crystals implies that the localized states involved in the conduction process may not be spin polarized through alignment of the Co ion magnetic moments.

Although the formation of localized electronic states owing to the incorporation of Co in ZnO is not controversial, our results support the idea that the electronic transport properties in this DMS may be only weakly related to the magnetic properties. Some results on magnetic tunnel junctions fabricated from ferromagnetic cobalt doped ZnO have non-saturating positive magnetoresistance at low temperatures~\cite{Song_both,Chen}, rather than negative magnetoresistance as would be expected from tunneling of spin-polarized charge carriers between layers with their magnetic moments aligned in high fields. Additionally, those authors acknowledged that the behavior of the tunneling magnetoresistance does not correlate with the magnetization of these devices. These results point to the possibility that the existence of ferromagnetism in cobalt doped ZnO is not a sufficient condition for generation of spin-polarized charge carriers.

\section{Conclusions}
Pristine and cobalt doped ZnO single crystals were grown by a molten salt solvent technique at low temperatures (480~$^0$C). The Zn$_{1-x}$Co$_x$O crystals are paramagnetic; there is no sign of ferromagnetism at up to $x=0.10$. The paramagnetic properties of the crystals arise from isolated Co$^{2+}$ ions.  At temperatures above 100~K, conduction in the crystals is by thermal activation, with a change to Mott variable range hopping at lower temperatures. Doped and undoped ZnO crystals show almost negligible negative MR at room temperature, with a crossover to positive MR below 20~K. The positive MR reaches values as large as 20\% in the Zn$_{1-x}$Co$_x$O crystals at 2.5~K, and is consistent with a field-induced shrinking of the localized state wavefunctions, which reduces the probability of a hop. The magnetotransport is decoupled from the magnetization, and no spin dependent transport is observed at the highest level of magnetic alignment of the Co$^{2+}$ ions. Our results indicate that the Co incorporated in low temperature grown single crystals simultaneously introduces paramagnetic moments and localized electronic states into the ZnO electronic structure, however the bulk magnetic and magnetotransport properties are independent of each other.

\begin{acknowledgments}
One of the authors, N.S., gratefully acknowledges support provided by the exchange agreement between the Indian Institute of Technology Delhi, India and the Ecole Polytechnique F\'{e}d\'{e}rale de Lausanne, Switzerland.
\end{acknowledgments}

\end{document}